\documentclass[twoside]{article}
\usepackage{fleqn,espcrc2,epsf}
\usepackage{graphicx}
\usepackage{epsfig}
\usepackage[figuresright]{rotating}

\newcommand{\be}[1]{\begin{equation} \label{(#1)}}
\newcommand{\ee}{\end{equation}}
\newcommand{\ba}[1]{\begin{eqnarray} \label{(#1)}}
\newcommand{\ea}{\end{eqnarray}}

\newcommand{\AmS}{{\protect\the\textfont2
  A\kern-.1667em\lower.5ex\hbox{M}\kern-.125emS}}

\def\PRD{{\em Phys. Rev.} {\bf D}}

\def\be{\begin{equation}}
\def\ee{\end{equation}}
\def\bea{\begin{eqnarray}}
\def\eea{\end{eqnarray}}
\def \KK {H.V.~Klapdor-Kleingrothaus}
\title{GENIUS - A New Facility of Non-Accelerator Particle Physics}

\author{H.V. Klapdor--Kleingrothaus
\address{Max--Planck--Institut f\"ur Kernphysik, 
P.O.Box 10 39 80, D--69029 Heidelberg, Germany\\
Spokesman of HEIDELBERG-MOSCOW and GENIUS Collaborations\\
e-mail:klapdor@gustav.mpi-hd.mpg.de, home page: http://mpi-hd.mpg.de.non-acc/}
}

\begin{document}

\begin{abstract}
	The GENIUS (\underline {Ge}rmanium in Liquid 
	\underline {Ni}trogen \underline {U}nderground \underline {S}etup) 
	project has been proposed in 1997 
\cite{KK-BEY97} 
	as first third generation double beta decay project, with 
	a sensitivity aiming down to a level of an effective neutrino 
	mass of 
$<m>\sim$ 0.01 - 0.001 eV. 
	Such sensitivity has been shown 
	to be indispensable to solve the question of the structure of the 
	neutrino mass matrix which cannot be solved by neutrino oscillation 
	experiments alone 
\cite{KKPS}. 
	It will allow broad access also to many other topics of 
	physics beyond the Standard Model of particle physics at the 
	multi-TeV scale. For search of cold dark matter GENIUS will cover 
	almost the full range of the parameter space of predictions of SUSY 
	for neutralinos as dark matter 
\cite{KK-Ram,Bed-KK2}. 
	Finally, GENIUS has the potential to be the first real-time detector 
	for low-energy (pp and $^7{Be}$) solar neutrinos 
\cite{Bau-KK,KKPropos99}. 
	A GENIUS-Test Facility has just been funded and will come into 
	operation by end of 2001.
\vspace{1pc}
\end{abstract}

\maketitle

\section{Introduction}

	Underground physics can complement in many ways the search for 
	New Physics at future colliders such as LHC and NLC and can serve 
	as important bridge between the physics that will be gleaned from 
	future high energy accelerators on the one hand, and satellite 
	experiments such as MAP and PLANCK on the other 
\cite{KK60Y}. 
	The first indication for beyond SM physics indeed has come from 
	underground experiments (neutrino oscillations from SK), and this 
	type of physics will play an even large role in the future.

	Concerning neutrino physics, without double beta decay there will be 
	no solution of the nature of the neutrino (Dirac or Majorana 
	particle) and of the structure of the neutrino mass matrix. Only 
	investigation of $\nu$ oscillations {\it and} double beta decay 
	together can lead to an absolute mass scale.

	Concerning the search for cold dark matter, even a discovery of 
	SUSY by LHC will not have proven that neutralinos form 
	indeed the cold dark matter in the Universe. Direct detection 
	of the latter by underground detectors remains indispensable. 
	Concerning solar neutrino physics, present information on possible 
$\nu$ 
	oscillations relies on 
0.2$\%$ 
	of the solar neutrino flux. The total pp neutrino flux has not 
	been measured and also no real-time information is available for 
	the latter.

	The GENIUS project proposed in 1997 (see 
\cite{KK-BEY97,KK60Y})
	as the first third generation $\beta\beta$ detector, could attack 
	all of these problems with an unpredented sensitivity.


\section{GENIUS, Double Beta Decay and the Light Majorana Neutrino Mass}

	Present double beta experiments are not able to reach a limit 
	for the (effective) neutrino mass below 
$\sim$ 0.1 eV. 
	The most sensitive experiment is {\it since eight years} the 
	HEIDELBERG-MOSCOW experiment using the world's largest source 
	strength of 11 kg of 86$\%$ enriched $^{76}{Ge}$ 
	in form of 5 high-purity Ge detectors, run in the Gran Sasso 
	Underground laboratory. The limits reached after 37.24 kg y 
	of measurement\\ 
	$T^{0\nu}_{1/2}$ $>3.5(2.1)\cdot{10}^{25}y$ and 
	$<m_{\nu}><0.26 (0.34) eV$, 68$\%$ and 90$\%$ c.l., respectively.

	The status and potential of other experiments are shown in 
Fig. 1.

\begin{figure}[htb]
\vspace{9pt}
\centering{
\includegraphics*[width=50mm, height=75mm, angle=-90]{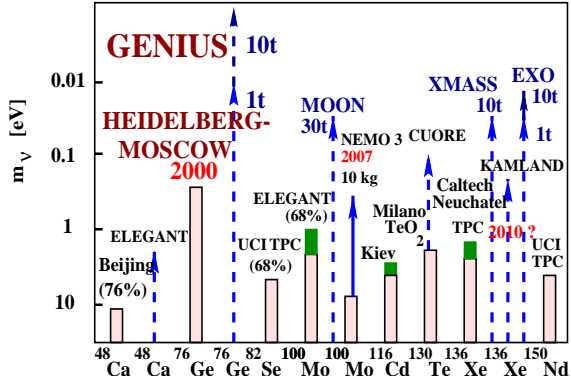}}
\caption{Present situation, 2000, and expectation for the future, of the 
	most promising $\beta\beta$ experiments. Light parts of the bars: 
	present status; dark parts: expectation for running experiments; 
	solid and dashed lines: experiments under construction  
	or proposed experiments. For references see \cite{LowNu2}.}
\end{figure}

	With the era of the HEIDELBERG-MOSCOW experiment which will 
	remain the most sensitive experiment for the next years, the time 
	of the small smart experiments is over.

\begin{figure}[htb]
\vspace{9pt}
\centering{
\includegraphics*[width=50mm, height=75mm, angle=-90]
{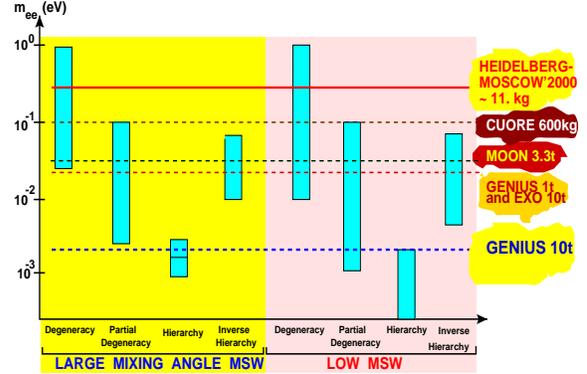}}
\caption{Values expected from $\nu$ oscillation experiments for 
	$m_{ee}\equiv(<m_\nu >)$ 
	in different schemes. The expectations are compared with the 
	present neutrino mass limits {\it obtained} from the 
	HEIDELBERG-MOSCOW experiment as well as the {\it expected}  
	sensitivities for the CUORE, MOON, EXO proposals and the 1 ton 
	and 10 ton proposal of GENIUS \cite{KKP}. For references  and more 
	details about the different experiments see 
\cite{KK60Y,LowNu2}.}
\end{figure}


The requirements in sensitivity for future experiments to play 
	a decisive role in the solution of the structure of the neutrino 
	mass matrix are shown in 
Fig.2.
	Shown are the expectations for the effective neutrino mass (the 
	observable in $\beta\beta$ decay) from the present experimental 
	status of all existing neutrino oscillation experiments in the 
	different presently experimentally favored neutrino mass models 
\cite{KKPS,KKP}.

	It can be seen that a sensitivity down to\\ 
$<m_{\nu}>\approx$ 0.001 eV as it may be reached {\it only} by the GENIUS 
	project will be able to test {\it all} neutrino scenarios 
	allowed by the oscillation experiments, except for one, the not 
	favoured hierarchical LOW solution. For details see
\cite{KKPS,KKP,KKPcomm}. 

	To reach this level of sensitivity $\beta\beta$ experiments have 
	to become large. A source strength of up to 10 tons of enriched 
	material touches the world production limits. At the same time the 
	background has to be reduced by a factor of 1000 and more compared 
	to that of the HEIDELBERG-MOSCOW experiment. 

Table 1 
	lists some key numbers for GENIUS, and of the main other proposals 
	made after the GENIUS proposal. Their potential is shown also 
in Fig.2. 
	It is seen that not all of these proposals fully cover the region 
	to be probed. Among them is also the recently presented MAJORANA 
	project. 
	

	In the GENIUS project a reduction by a factor of more than 1000 down 
	to a background level of 0.1 events/tonne y keV in the range of 
$0\nu\beta\beta$ is reached by removing all material close to the detectors, 
	and by using naked Germanium detectors in a large 
	tank of liquid nitrogen. It has been shown that the detectors show 
	excellent performance under such conditions
\cite{KKPropos99}.
	
	For technical questions and extensive Monte Carlo simulations of the 
	GENIUS project for its application in double beta decay we refer to 
\cite{KKPropos99}.

\vspace{190pt}
\begin{figure}[t!]
\begin{picture}(100,145)
\centering{
\put(0,-120){\includegraphics{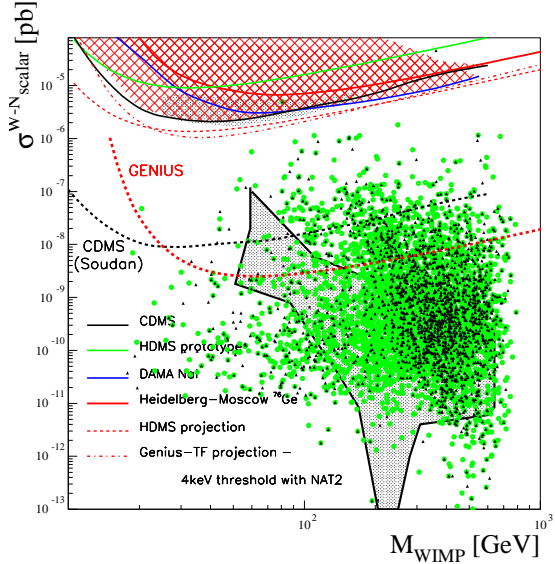}}}
\end{picture}

\vspace{30pt}
\caption{WIMP-nucleon cross section limits in pb for scalar
	interactions as function of the WIMP mass in GeV.
	Shown are contour lines of present experimental limits (solid lines) 
	and of projected experiments (dashed lines). Also shown is the 
	region of evidence published by DAMA. The theoretical expectations 
	are shown by a scatter plot (from \cite{Bed-KK2}) and by the grey 
	region (from \cite{Ell}).} 
\end{figure}


\section{GENIUS and Other Beyond Standard Model Physics}

	GENIUS will allow besides the major step in neutrino physics 
	described in section 2 the access to a broad range of other beyond 
	SM physics topics in the multi-TeV range. Already now 
	$\beta\beta$ decay probes the TeV scale on 
	which new physics should manifest itself (see, e.g. 
\cite{KK-BEY97,KK-TR98}). 
	Basing to a large extent on the theoretical work of the Heidelberg 
	group in the last four years, the HEIDELBERG-MOSCOW experiment yields 
	results for SUSY models (R-parity breaking, sneutrino mass), 
	leptoquarks (leptoquarks-Higgs coupling), compositeness, 
	right-handed W mass, nonconservation of Lorentz invariance and 
	equivalence 
	principle, mass of a heavy left or righthanded neutrino, 
	competitive to corresponding results from high-energy accelerators 
	like TEVATRON and HERA. The potential of GENIUS extends into the 
	multi-TeV region for these fields and its sensitivity would 
	correspond to that of LHC or NLC and beyond (for details see 
\cite{KK60Y,KK-TR98}).


\section{GENIUS and Cold Dark Matter Search}

	Already now the HEIDELBERG-MOSCOW experiment is the most sensitive 
	Dark Matter experiment worldwide concerning the raw data. GENIUS 
	would already in a first step, with 100 kg of {\it natural} Ge 
	detectors, cover a significant part of the MSSM parameter space 
	for prediction of neutralinos as cold dark matter 
(Fig. 3). For this purpose the background 
	in the energy range $<$ 100 keV has to be reduced to 
	${10}^{-2}$ events/kg y keV, which is possible if the detectors 
	are produced and handled on Earth surface under heavy shielding, 
	to reduce the cosmogenic background produced by spallation through 
	cosmic radiation (critical products are tritium, 
$^{68}{Ge}$, $^{63}{Ni}$, ...) 
	to a minimum. For details we refer to 
\cite{KKPropos99}. 
Fig. 3 shows together with the expected sensitivity of GENIUS predictions 
	for neutralinos as dark matter by two models, one basing on 
	supergravity
\cite{Ell}, 
	another starting from more relaxed unification conditions  
\cite{Bed-KK2}.
 
	The sensitivity of GENIUS for Dark Matter corresponds to that 
	obtainable with a 1 ${km}^3$ AMANDA detector for 
	{\it indirect} detection (neutrinos 
	from neutralino annihilation at the Sun). Interestingly both 
	experiments would probe different neutralino 
	compositions: GENIUS mainly gaugino-dominated neutralinos, 
	AMANDA mainly neutralinos with comparable 
	gaugino and Higgsino components. 
	It should be stressed that, together with DAMA, GENIUS will be the 
	only future Dark Matter experiment, which would be able to positively 
	identify a dark matter signal by the seasonal modulation signature. 
	This {\it cannot} be achieved, for example, by the CDMS experiment.

\begin{table*}[h]
\caption{Some key numbers of future double beta decay experiments (and of 
	the {\sf HEIDELBERG-MOSCOW} experiment). Explanations: 
	${\nabla}$ - assuming the background of the present pilot project. 
	$\ast\ast$ - with matrix element from [Sta90*-II], [Tom91**-I], 
	[Hax84**-I], [Wu91*-II], [Wu92*-II] (see Table II in [HM99*-III]). 
	${\triangle}$ - this case shown 
	to demonstrate {\bf the ultimate limit} of such experiments. 
	For details see \cite{KK60Y}.}
\label{table:1}
\newcommand{\m}{\hphantom{$-$}}
\newcommand{\cc}[1]{\multicolumn{1}{c}{#1}}
\renewcommand{\tabcolsep}{0.9pc} 
\renewcommand{\arraystretch}{1.} 
{\footnotesize
{  
\begin{tabular}[!h]{|c|c|c|c|c|c|c|c|}
\hline
\hline
 &  &  &  & Assumed &  &  & \\
 &  &  &  & backgr. & Run- & Results & \\
$\beta\beta$-- & & & Mass & $\dag$ events/ & $ning$ & limit for & 
${<}m_{\nu}{>}$ \\
$Isoto-$ & $Name$ & $Status$ & $(ton-$ & kg y keV, & Time  
& $0\nu\beta\beta$ & \\
pe & & & nes) & $\ddag$ events/kg & (tonn. & half-life & ( eV )\\ 
& & & & y FWHM,  & years) & (years) & \\
& & & & $\ast$ events &  &  & \\
& & &  & /yFWHM &  &  & \\
\hline
\hline
 &  &  &  &  &  &  & \\
~${\bf ^{76}{Ge}}$ & {\bf HEIDEL-} & {\bf run-}  & 0.011 & $\dag$ 0.06 
& {\bf 35.5} & ${\bf 1.9\cdot{10}^{25}}$ & {\bf $<$ 0.34} $\ast\ast$\\
 & {\bf BERG}  &  &  (enri-  &  &  {\bf kg y} &  {\bf 90$\%$ c.l.} 
& {\bf 90$\%$ c.l.} \\
& {\bf MOSCOW} & {\bf ning} & ched) & $\ddag$ 0.24  &  
& ${\bf 3.1\cdot{10}^{25}}$ & {\bf $<$ 0.26} $\ast\ast$\\
& {\bf [Kla99e**]} &  &  & $\ast$ 2 & & 
{\bf 68$\%$ c.l.} & {\bf 68$\%$ c.l.}\\
& {\bf [HM2000*]} &  &  &  &  & {\bf NOW !!}  & {\bf NOW !!}\\
& {\bf [-III]} &  &  &  &  &  &\\
\hline
\hline
\hline
 &  &  &  &  &  &  & \\
${\bf ^{100}{Mo}}$ & {\sf NEMO III} & {\it under} & $\sim$0.01 & $\dag$ 
{\bf 0.0005} &  &  &\\
 & {\tt [NEM2000]}& {\it constr.} & (enri- & $\ddag$ 0.2  & 50 & 
${10}^{24-25}$ & 0.3-0.7\\
 &  &  & -ched) &  $\ast$ 2 &kg y  &  &\\
\hline
\hline
&  &  &  &  &  &  & \\
${\bf ^{130}{Te}}$ & ${\sf CUORE}^{\nabla}$ & Pro- & 0.75 & $\dag$ 0.5 & 5 & 
$9\cdot{10}^{24}$ & 0.2-0.5\\
 & {\tt [Gui99a*}& posal &(natu-  & $\ddag$ 4.5  &  & & \\ 
 &{\tt  -VI]} &  & ral)& $\ast$ 1000  &  &  & \\
\hline
&  &  &  &  &  &  &  \\
${\bf ^{130}{Te}}$ & {\sf CUORE}  &  Pro- & 0.75   & $\dag$ 0.005 & 5 
& $9\cdot{10}^{25}$ & 0.07-0.2\\
&  &  posal  & (natu-  &  $\ddag$ 0.045&  & &\\
&  & &  ral) &  $\ast$ 45 &  &  & \\
\hline
&  &  &  &  &  &  & \\
${\bf ^{100}{Mo}}$ & {\sf MOON} & idea & 10 (en-& ? & 30 & ? & \\
 & {\tt [Eji99b*} &  & rich.) & & & &0.03 \\
 & {\tt -VI]} &  & 100 &  & 300 &  & \\
 &  &  & (nat.) &  &  &  & \\
\hline
&  &  &  &  &  &  & \\
${\bf ^{136}{Xe}}$ & {\sf EXO} & Pro-& 1 & $\ast$ 0.4 & 5 & 
$8.3\cdot{10}^{26}$ & 0.05-0.14\\
&  & &  &  &  &  & \\
${\bf ^{136}{Xe}}$ & {\tt [Dan2000a]} & posal  & 10 & $\ast$ 0.6 & 10 & 
$1.3\cdot{10}^{28}$ & 0.01-0.04\\
&  &  &  &  &  &  & \\ 
\hline 
\hline
\hline
\hline
&  &  &  &  &  &  &  \\
~${\bf ^{76}{Ge}}$ & {\bf GENIUS} & Pro- & 1  & $\dag$ 
${\bf 0.04\cdot{10}^{-3}}$ & 1 & ${\bf 5.8\cdot{10}^{27}}$ & 
{\bf 0.02-0.05} \\
 & {\tt [Kla97**}  & posal &(enrich.)  
& $\ddag$ ${\bf 0.15\cdot{10}^{-3}}$ & & & \\
& {\tt -VI]} &  &  & $\ast$ {\bf 0.15} &  &  &  \\
&  &  & 1 & ${\bf \ast~ 1.5}$ & 10 & ${\bf 2\cdot{10}^{28}}$  & 
{\bf 0.01-0.028} \\
\hline
&  &  &  &  &  &  &  \\
~${\bf ^{76}{Ge}}$ & {\bf GENIUS} & Pro- & 10 
& $\ddag$ ${\bf 0.15\cdot{10}^{-3}}$ & 10 &
${\bf 6\cdot{10}^{28}}$ & {\bf 0.006 -}\\
&  {\tt [Kla97**-} &  &  &  &  &  &  {\bf 0.016}\\
 &  {\tt -VI]} &  posal &(enrich.) & ${\bf 0^{\triangle}}$ & 10 & 
${\bf 5.7\cdot{10}^{29}}$ & {\bf 0.002 -}\\
&  &  &  &  &  &  &  {\bf 0.0056}\\ 
\hline 
\hline
\end{tabular}\\[2pt]
}}
\end{table*}


\begin{figure}[htb]
\vspace{9pt}
\centering{
\includegraphics*[width=60mm, height=80mm, angle=-90]{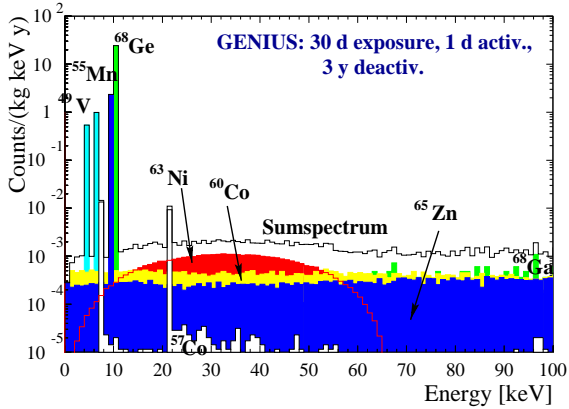}} 
\caption{Simulated cosmogenic background during detector production. 
	Assumptions: 30 days exposure of material before processing, 
	1 d activation after zone refining, 3 y deactivation underground 
	(see \cite{LowNu2}).}
\end{figure}



\section{GENIUS and Low-Energy Solar Neutrinos}

	Gallex and Sage 
	measure pp + $^7{Be}$ + $^{8}B$ neutrinos (60 + 30 + 10$\%$) down 
	to 0.24 MeV, the Chlorine experiment measured $^7{Be}$ + $^8{B}$ 
	neutrinos (80$\%$ $^8{B}$) above $E_\nu$= 0.817 MeV, all without 
	spectral, time and detection information. No experiment has 
	separately measured the pp and $^7{Be}$ neutrinos and no experiment 
	has measured the {\it full} pp $\nu$ flux. BOREXINO plans to 
	measure $^7{Be}$ neutrinos, the access to pp neutrinos 
	being limited by $^{14}C$ contamination (the usual problem of 
	organic scintillators). GENIUS could be the first detector measuring 
	the {\it full} pp ( and $^7{Be}$) neutrino flux in real time.

\begin{figure}[htb]
\vspace{9pt}
\centering{
\includegraphics*[width=80mm, height=55mm]{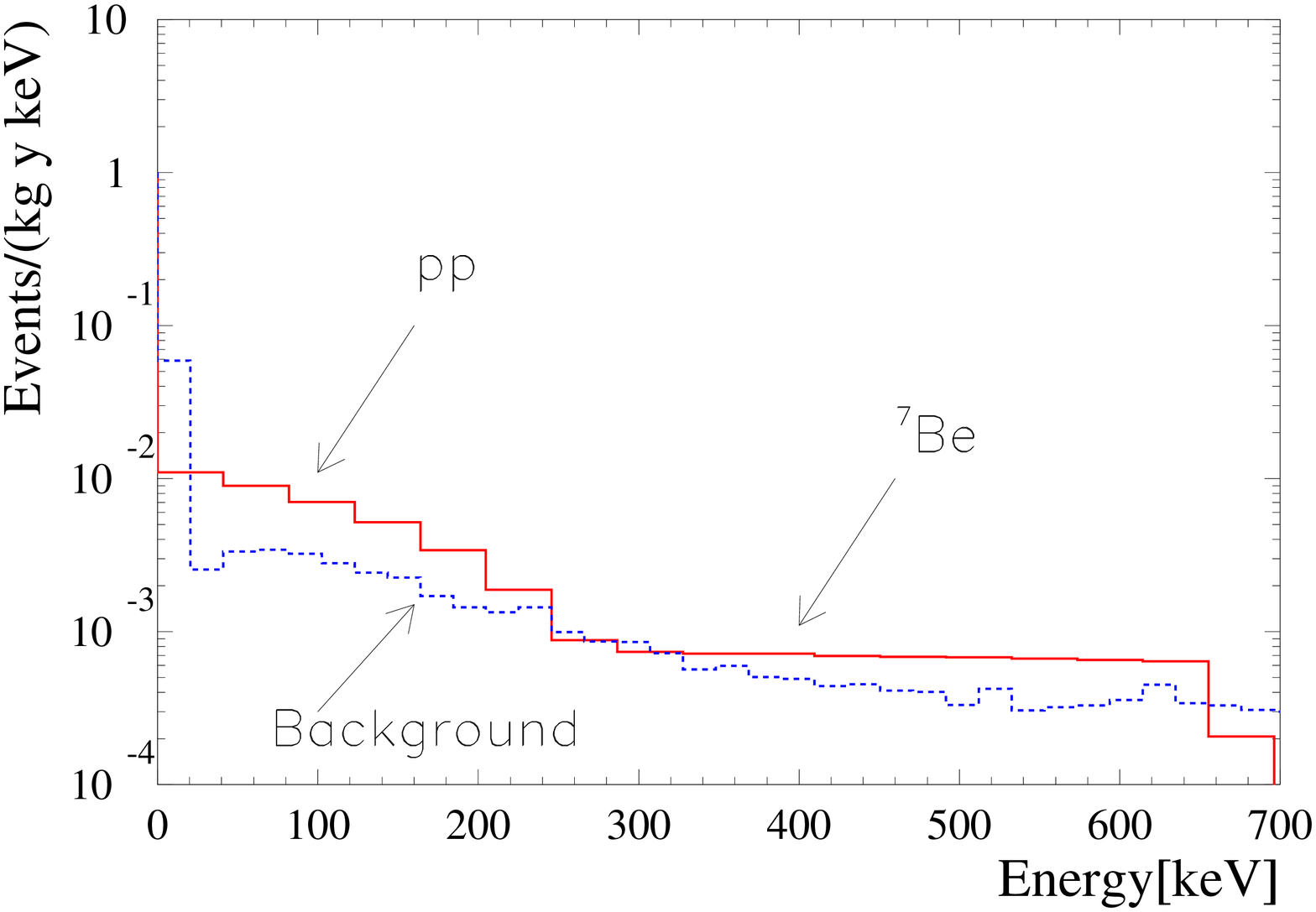}}
\caption{Simulated spectrum of low-energy solar neutrinos (according to SSM) 
	for the GENIUS detector (1 tonne natural or enriched Ge) 
(from \cite{Bau-KK}).}
\end{figure}

	Extending the radius of GENIUS to 13 m and improving some of the 
	shielding parameters as described in 
\cite{KKPropos99,Bau-KK} 
	the background can be reduced to a level of 
${10}^{-3}$ events/ kg y keV (Fig. 4) (see also 
\cite{LowNu2}). 
	This will allow to look for the pp and $^7{Be}$ solar neutrinos by 
	elastic neutrino-electron scattering with a threshold of 11 keV or at 
	most 19 keV (limit of possible tritium background) (Fig. 5) which 
	would be the 
	lowest threshold among other proposals to detect pp 
	neutrinos, such as HERON, HELLAZ, NEON, LENS, MOON, XMASS.

	The counting rate of GENIUS (10 ton) would be 6 events per day 
	for pp and 18 per day for $^7{Be}$ neutrinos, i.e. similar to 
	BOREXINO, but by a factor of 30 to 60 larger than a 20 ton LENS 
	detector and a factor of 10 larger than the MOON detector.


\section{GENIUS - Test Facility}

	Construction of a test facility for GENIUS - GENIUS-TF - 
	consisting of $\sim$ 40 kg of HP Ge detectors suspended in a 
	liquid nitrogen box has been started. Up to end of 2000, three 
	detectors each of $\sim$ 2.5 kg and with a threshold of as low as 
	$\sim$ 500 eV have been produced.

	Besides test of various parameters of the GENIUS project, the test 
	facility would allow, with the projected background of 
	4 events/kg y keV in the low-energy range, to probe the DAMA evidence 
	for dark matter by the seasonal modulation signature within about 
	one year of measurement  
	with 95 $\%$ c.l.. Even for an initial lower mass of 20 kg the 
	time scale would be not larger than three years (for details see 
\cite{KK2000,Bau2000}. 
	If using the enriched $^{76}{Ge}$ detectors of the HEIDELBERG-MOSCOW 
	experiment in the GENIUS-TF setup, a background in the 
	$0\nu\beta\beta$ region a factor 30 smaller than in the 
	HEIDELBERG-MOSCOW experiment could be obtained, 
	which would allow to test the effective Majorana neutrino mass down 
	to 0.15 eV (90$\%$ c.l.) in 6 years of measurement. This limit is 
	similar to what much larger experiments aim at, at much larger 
	time scale (see Table 1.).


\section{Conclusion}

	The GENIUS project is - among the projected or discussed other third 
	generation double beta detectors - the one which exploits this method 
	to obtain information on the neutrino mass to the ultimate limit. 
	Nature is extremely generous to us, that with an increase of the 
	sensitivity by two orders of magnitude compared to the present 
	limit, down to $<m_\nu>\sim {10}^{-3}$ eV, indeed essentially all 
	neutrino scenarios allowed by present neutrino oscillation 
	experiments can be probed.

	GENIUS is the only of the new projects which 
	simultaneously has a huge potential for cold dark matter search, 
	{\it and} for real-time detection of low-energy neutrinos. 


\end{document}